\documentclass[aps,prl,superscriptaddress,showpacs,floatfix]{revtex4-1}
\usepackage{graphicx,amsmath,amssymb,xspace,epsfig,float,multirow,subfigure,tabularx}
\usepackage{amsbsy,amssymb,amsmath,bm}
\usepackage{epsf}
\usepackage{color,natbib}
\usepackage{amsfonts}
\usepackage{hyperref}

\pdfoutput=1

\begin{document}

\title{Aslamazov Larkin Conductivity in Weyl Semimetals}
\author{B.Ya.Shapiro}
\affiliation{Department of Physics, Institute of Superconductivity, Bar Ilan University, Ramat Gan, 52100,Israel}
\email{shapib@biu.ac.il}

\begin{abstract}
The Aslamazov-Larkin conductivity (ALC) in the Weyl semi-metals is
calculated both for Type-I and Type II for different dimensionality and
magnetic fields. The ALC strongly depends on the tilt parameter of the
dispersion relation cone. While the 3D and 2D the Aslamazov -Larkin
conductivity slightly depends on tilt parameter in Type-I phase and
increases in the Type-II phase the 1D ALC decreases in Type-I phase up to
zero close to border between Type-I and Type-II phases. Results are
discussed in light of the resent experiments on the layered $HfTe_{5}$ Weyl
semi-metals. It is concluded the one dimensional AL conductivity well
explained the experimental data.
\end{abstract}

\maketitle

\affiliation{Department of Physics and Institute of Superconductivity, Bar
Ilan University, Ramat Gan, Bar Ilan, 52100,Israel}

\affiliation{Department of Physics and Institute of Superconductivity, Bar
Ilan University, Ramat Gan, Israel, 52100}

\affiliation{Department of Physics and Institut of Superconductivity, Bar
Ilan University, Ramat Gan, Israel, 52100}

\affiliation{Department of Physics and Institute of Superconductivity, Bar
Ilan University, Ramat Gan, Israel, 52100}

\affiliation{Department of Physics and Institute of Superconductivity, Bar
Ilan University, Israel, Ramat Gan, 52100}

\affiliation {Department of Physics, Institute of Superconductivity, Bar-Ilan
University, Ramat-Gan 52900, Israel}

\affiliation{Physics Department, Bar-Ilan University, 52900 Ramat-Gan,
Israel}

\section{Introduction.}

Dispersion relation near Fermi surface in recently synthesized two and three
dimensional Weyl (Dirac) (WSM/DSM) semi-metals \cite{one},\cite{two},\cite%
{three} is qualitatively distinct from conventional metals, semi - metals or
semiconductors in which all the bands are parabolic. In type I Weyl
semi-metals (WSM), the band inversion results in Weyl points in low-energy
excitations being anisotropic massless "relativistic" fermions with
dispersion relation of these semimetals looks as a tilted Dirac cone. In
Type I WSM/DSM the tilt angle of the cone is smaller than some critical
value \cite{four} while recently discovered layered transition-metal
dichalcogenides is defined as type-II WSM/DSM \cite{five}. \cite{six}. In
this latter case, the Dirac cone exhibits a strong tilt, so that it can be
characterized by a nearly flat band at Fermi surface. The type-II WSM also
exhibit exotic topological properties different from the type-I ones, such
anti-chiral effect of the chiral Landau level and novel quantum
oscillations. Graphene is a prime example of the type I WSM, while
materials, like layered organic compound were long suspected to be a 2D
type-II Dirac fermions. Several materials were observed to undergo the type-
I to II transition while doping or pressure is changed \cite{seven},\cite%
{eigth},\cite{nine}. Theoretically physics of the topological (Lifshitz)
phase transitions between the type I to type II Weyl semi-metals were
considered in the context of superfluid phase in $He_{3}$ , layered organic
materials in 2D and 3D Weyl semi-metals \cite{ten},\cite{elleven},\cite%
{twelve}. The pressure modifies the spin orbit coupling that in turn
determines the topology of the Fermi surface of these novel materials \cite%
{13}.

Many Weyl and Dirac materials are known to be superconducting. A detailed
study of superconductivity in WSM/DSM under hydrostatic pressure revealed a
curious dependence of critical temperature of the superconducting transition
on pressure. In particular the critical temperature in some of these systems
like $HfTe_{5}$ show a sharp maximum as a function of pressure \cite{14}.
This contrasts with generally smooth dependence on pressure in other
superconductors (not suspected to be Weyl materials) like cuprates. Magnetic
properties in these topological superconductors are also very different from
that in conventional superconductors. In particular the Abrikosov parameter
used to distinguish between the superconductivity of the first from the
second type depends on the cone tilt and may totally change magnetic
properties varying from first kind superconductor (like clean metals) to the
second kind. The critical fields, coherence lengths magnetic penetration
depths and the Ginzburg number characterizing the strength of fluctuations
strongly depend on cone tilt \cite{15,16}. It reveals an extremely important
relation between the cone title and fluctuation in WSM/DSM superconductors 
\cite{17}. We begin with a brief review of studies of fluctuations in
superconductors. The subject was initiated in the work of Aslamazov and
Larkin \cite{18} calculated the conductivity of fluctuating Cooper pairs in
zero magnetic field. Maki and Thomson \cite{19,20,21} included effects of
electron scattering off the fluctuations. It was found that there is another
badly divergent contribution known as anomalous Maki-Thomson correction.
Physically, this correction is connected with the coherent scattering of the
electrons by the impurities and analogous to the weak localization
correction. The divergence can be removed by introducing a pair-breaking
rate. It should be noted that, the experimental results at temperature close
to the critical $T_{c0}$ can be described by the Aslamazov-Larkin term only.
This suggests that the pair-breaking rate is relatively large in real
superconductors. Later, Thomson and Maki returned to the issue and evaluated
fluctuation correction to the normal conductivity in finite fields \cite{22}%
, \cite{23}. A theory of transport phenomena in the fluctuation region in
the dirty, clean and super clean limits was developed by Aronov et al. \cite%
{24} . Their consideration was based on the Ginzburg-Landau equations and,
thus, is applicable for relatively small fields. Experimentally the
broadened of the normal-to-superconductivity transition curve caused by the
superconducting fluctuations was studied in a lot of WSM/DSM materials while
the 2D Aslamazov Larkin conductivity was detected in 3D WSM material $%
Cd_{2}As_{3}$ \cite{25}. In the present paper we study of Aslamazov-Larkin
conductivity (ALC) in Weyl/Dirac semimetals. The focus generally is on the
dependence of the ALC in the cone tilt parameter and consequently on the ALC
in the vicinity of the transition from type-I to type-II WSM. 
\begin{figure}[h]
\centering \includegraphics[width=10cm]{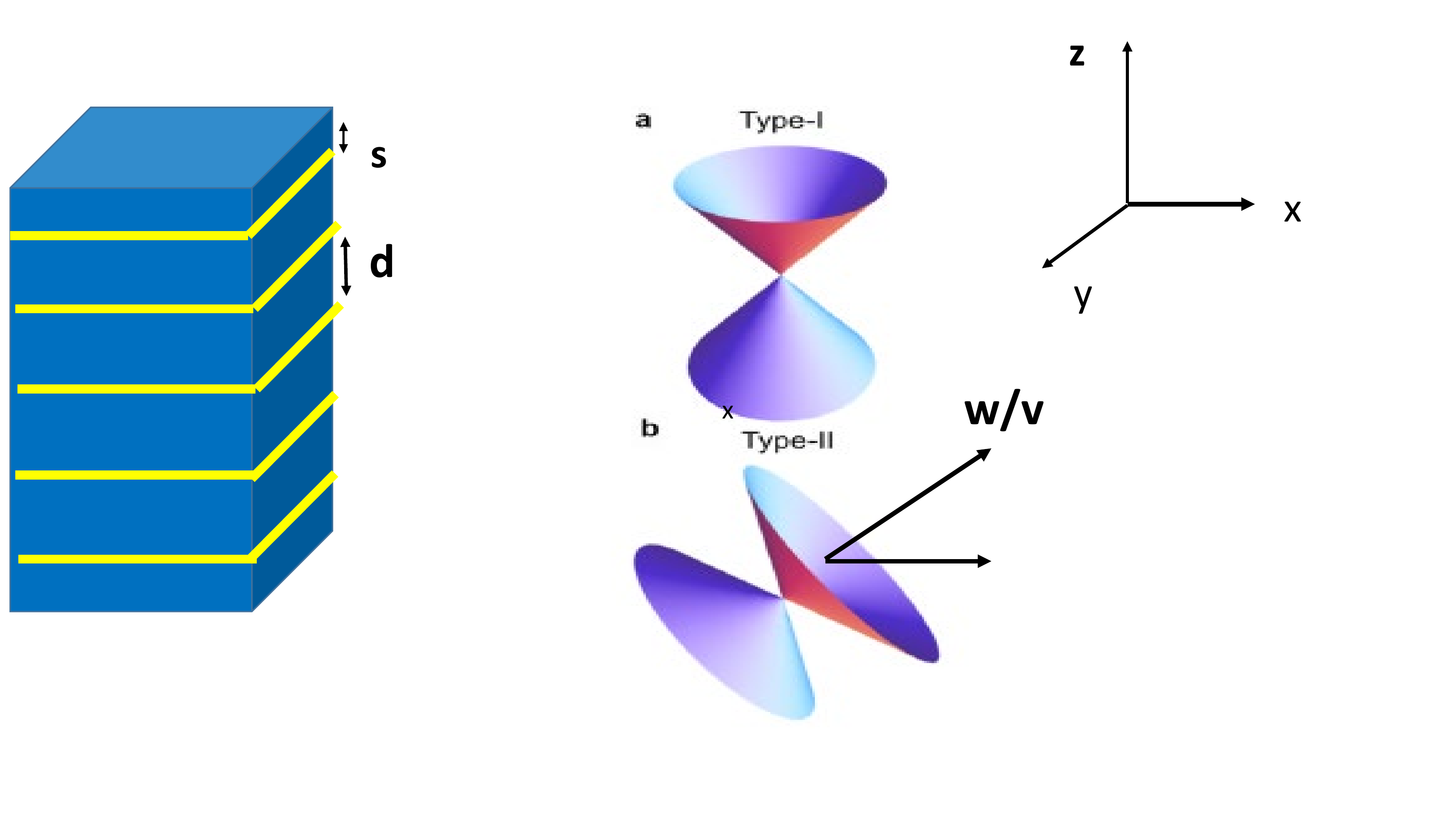}
\caption{ Geometry of the problem}
\end{figure}

\section{Model}

Since the Weyl material typically possesses several sublattices the effect
of the topological transition on superconductivity was exemplified using the
simplest possible model with just two sublattices. The band structure near
the Fermi level of a 3D layered Weyl semi-metal is well captured by the
non-interacting massless Weyl dispersion relation with the "in plane" Fermi
velocity $\mathbf{v}$ (assumed to be isotropic in the $x-y$ plane) and
conventional parabolic term on $z-$direction\cite{26},\cite{27},\cite{28},%
\cite{29} (see Fig.1) and described by the Hamiltonian

\begin{eqnarray}
K &=&\int_{\mathbf{r}}\psi _{\alpha }^{s+}\left( \mathbf{r}\right) \widehat{K%
}_{\alpha \beta }\psi _{\beta }^{s}\left( \mathbf{r}\right) \text{\ \ \ \ }
\label{eq1} \\
\text{\ \ }\widehat{K}_{\gamma \delta } &=&-i\hbar v\nabla ^{i}\sigma
_{\gamma \delta }^{i}+\left( -i\hbar w_{i}\nabla ^{i}-\mu +\frac{p_{z}^{2}}{%
2m_{z}}\right) \delta _{\gamma \delta }\text{.}  \notag
\end{eqnarray}%
Here $\mu $ is the chemical potential, $p_{z}=-i\hbar \nabla _{z}$ , $\sigma 
$ are Pauli matrices in the sublattice in the sublattice space in the WSM
layers, with just two sublattices denoted by $\alpha =1,2$ and $s$ is spin
projection. The velocity vector $\mathbf{w}$ defines the tilt of the Dirac
dispersion cone. The graphene - like dispersion relation for $\mathbf{w}=0$
represents the type I Weyl semi-metal, while for the velocity $\left\vert 
\mathbf{w}\right\vert =w$ exceeding $v$, the material becomes a type II Weyl
semi - metal.

We restrict our self to the case of just one left handed and one right
handed Dirac points, typically but not always separated in the Brillouin
zone. Generalization to include the opposite chirality and several "cones"
is straightforward. We assume that different valleys are paired
independently and drop the valley indices.

\section{Ginzburg-Landau theory.}

\bigskip

The effective electron-electron attraction due to the electron - phonon
attraction opposed by Coulomb repulsion (pseudopotential) mechanism creates
pairing below $T_{c}$. Assuming the singlet s-channel interaction with
essentially local interaction in the layers one obtains the set of Gor'kov
equations \cite{27} and their microscopic Ginzburg-Landau expansion \cite{28}%
, \cite{29} for arbitrary Dirac cone tilt parameter $\kappa $ in the form:

\begin{equation}
F=\int d^{3}r\left\{ D_{0}\left( \mu \right) f\left( \kappa \right) \left(
\xi _{i}^{2}\left( \kappa \right) \left\vert \partial _{i}\Delta \right\vert
^{2}-\tau \left( \kappa \right) \left\vert \Delta \right\vert ^{2}+\frac{%
\beta \left( \kappa \right) }{2}\left\vert \Delta \right\vert ^{4}\right) +%
\frac{\left( \nabla \times \mathbf{A}\right) ^{2}}{8\pi }\right\} \text{,}
\label{Eq.2}
\end{equation}

\bigskip

Where $\tau \left( \kappa \right) =1-T/T_{c}\left( \kappa \right) ,\mathbf{%
\partial }_{i}=-i\hbar \mathbf{\nabla }_{i}-2eA/c,T_{c}\left( \kappa \right)
=1.14\Omega \exp \left( -1/\lambda _{0}f\left( \kappa \right) \right) .$

The index $i=x,y,z$ , $D_{0}\left( \mu \right) =\frac{\sqrt{2m_{z}}\mu ^{3/2}%
}{12\pi ^{2}\hbar ^{3}v^{2}},$ is the DOS for layered graphene $(\kappa =0)$%
, dimensionless constant $\lambda _{0}$ is the electron-electron strength
for zero tilt parameter $\kappa .$ $($We use below units with $c=\hbar =1)$.
The tunneling of the electrons moving between the superconducting layers via
dielectric streak described by the effective mass $m_{z}$ of the electrons
moving along the z axis. Within tight binding model the effective mass is
estimated as%
\begin{equation}
m_{z}=\frac{m_{e}s%
{{}^2}%
exp(d/s)}{d^{2}}  \label{mz}
\end{equation}
where $m_{e}$ is the mass of free electron, $d$ is the distance between
layers of thickness $s$, see Fig.1. The GL coefficients in Eq.\ref{Eq.2}
have been calculated for all values of cone tilt parameter $\kappa $ in our
Ref.\cite{29}. In particular for the type I WSM, $\kappa <1$, in which the
Fermi surface is a closed ellipsoid and it is given by:%
\begin{equation}
f\left( \kappa \right) =\frac{1}{\left( 1-\kappa ^{2}\right) ^{3/2}}\text{.}
\label{fI}
\end{equation}%
In the type II phase, $\kappa >1$, the Fermi surface becomes open, extending
over the Brillouin zone, and the corresponding expression is:%
\begin{equation}
f=\frac{\kappa ^{2}}{\ \pi \left( \kappa ^{2}-1\right) ^{3/2}}\left\{ 2\sqrt{%
1+\kappa }-1+\log \left[ \frac{2\left( \kappa ^{2}-1\right) }{\kappa \left(
1+\sqrt{1+\kappa }\right) ^{2}\epsilon }\right] \right\} \text{.}
\label{fII}
\end{equation}%
Here $\epsilon $ is an ultraviolet cut off parameter $\epsilon =$ $a\Omega
/\pi w$ , where $a$ is an interatomic spacing, $\Omega $ is the phonon
frequency around the Fermi energy.\bigskip

The critical temperature $T_{c}\left( \kappa \right) $ has the sharp spike
at the border between Type-I and Type-II states at $\kappa \rightarrow 1$
where the topology of the Fermi surface undergoes the Lifshitz 2.5 type
transition \cite{28}.

\section{Aslamazov-Larkin Conductivity.}

Using the standard time dependent Ginzburg Landau equations \cite{30} with
the external $\delta $ correlated random force responsible for thermal
fluctuations:

\begin{equation}
\Gamma \left( \frac{\partial }{\partial t}+2ie\Phi \right) \Delta
=D_{0}\left( \mu \right) f\left( \kappa \right) \left( \xi _{i}\left( \kappa
\right) \xi _{j}\left( \kappa \right) \partial _{i}\partial _{j}-\tau \left(
\kappa \right) +\beta \left( \kappa \right) \left\vert \Delta \right\vert
^{2}\right) \Delta +\Gamma T\zeta \left( r,t\right)  \label{eq3}
\end{equation}

\begin{equation}
\mathbf{J}=2eiD_{0}\left( \mu \right) f\left( \kappa \right) \xi
_{i}^{2}\Delta \left( \mathbf{r}\right) \mathbf{\partial }\Delta ^{\ast
}\left( \mathbf{r}\right) +c.c.+\sigma _{n}\left( -\frac{\partial \mathbf{A}%
}{\partial t}-\mathbf{\nabla }\Phi \right) \text{.}  \label{j}
\end{equation}%
where $\Gamma =D_{0}\left( \mu \right) f\left( \kappa \right) \pi
/8T_{c}\left( \kappa \right) $ is the relaxation constant $\Phi $ is the
electric potential, $\mathbf{A}$ is the vector potential, $\sigma _{n}$ is
the normal conductivity.

Taking order parameter in the form $\Delta (\mathbf{r},t)=\Delta _{0}(%
\mathbf{r})+\Delta _{1}(\mathbf{r},t)$ where $\Delta _{0}(\mathbf{r})$ is
the order parameter in the absence of the electric field, and substituting
it into Eqs.\ref{eq3},\ref{j} one obtains in the case of zero magnetic field
for the dc superconducting current:

\bigskip 
\begin{equation}
\mathbf{J}_{\alpha }=4\Gamma e^{2}i\xi _{i}^{2}\left( \kappa \right) \int_{%
\mathbf{k}}\frac{k_{\alpha }}{\xi _{s}^{2}\left( \kappa \right)
k_{s}^{2}+\tau }E_{\beta }\frac{\partial }{\partial \mathbf{k}_{\beta }}%
\left\langle \left\vert \Delta _{0}(\mathbf{k})\right\vert ^{2}\right\rangle
\   \label{jfluc}
\end{equation}

\bigskip where 
\begin{equation}
\left\vert \Delta _{0}(k\mathbf{,}\kappa )\right\vert ^{2}=\frac{T_{c}\left(
\kappa \right) }{D_{0}\left( \mu \right) f\left( \kappa \right) \left( \xi
_{s}^{2}\left( \kappa \right) k_{s}^{2}+\tau \left( \kappa \right) \right) }
\label{delta}
\end{equation}

\bigskip

Using \ref{jfluc} one obtains for the Aslamazov-Larkin conductivity: 
\begin{equation}
\sigma _{\alpha \alpha }=e^{2}\pi \xi _{\alpha }^{4}\left( \kappa \right)
\int_{\mathbf{k}}\frac{k_{\alpha }^{2}}{\left( \xi _{s}^{2}\left( \kappa
\right) k_{s}^{2}+\tau \left( \kappa \right) \right) ^{3}}  \label{sigma}
\end{equation}

Performing integration one obtains for different dimensionality. For 3D case
the tilt and temperature dependence of the ALC has the form:

\bigskip\ 
\begin{equation}
\sigma _{\alpha \alpha }\left( \kappa \right) =\frac{e^{2}\xi _{\alpha
}\left( \kappa \right) }{64\pi \xi _{\beta }\left( \kappa \right) \xi
_{\gamma }\left( \kappa \right) \tau \left( \kappa \right) ^{1/2}}\text{ }%
,\sigma _{zz}\left( \kappa \right) =\frac{e^{2}\xi _{z}\left( \kappa \right) 
}{32\xi _{x}\left( \kappa \right) \xi _{y}\left( \kappa \right) \tau \left(
\kappa \right) ^{1/2}}  \label{3D}
\end{equation}%
Where $\alpha =x,y.$

The reduced ALC $\sigma _{xx}=\sigma _{xx}\left( \kappa \right) /\sigma
_{xx}\left( 0\right) $ is shown in Fig.2 and demonstrates strong dependence
on the tilt parameter in Type-II phase. 
\begin{figure}[h]
\centering \includegraphics[width=10cm]{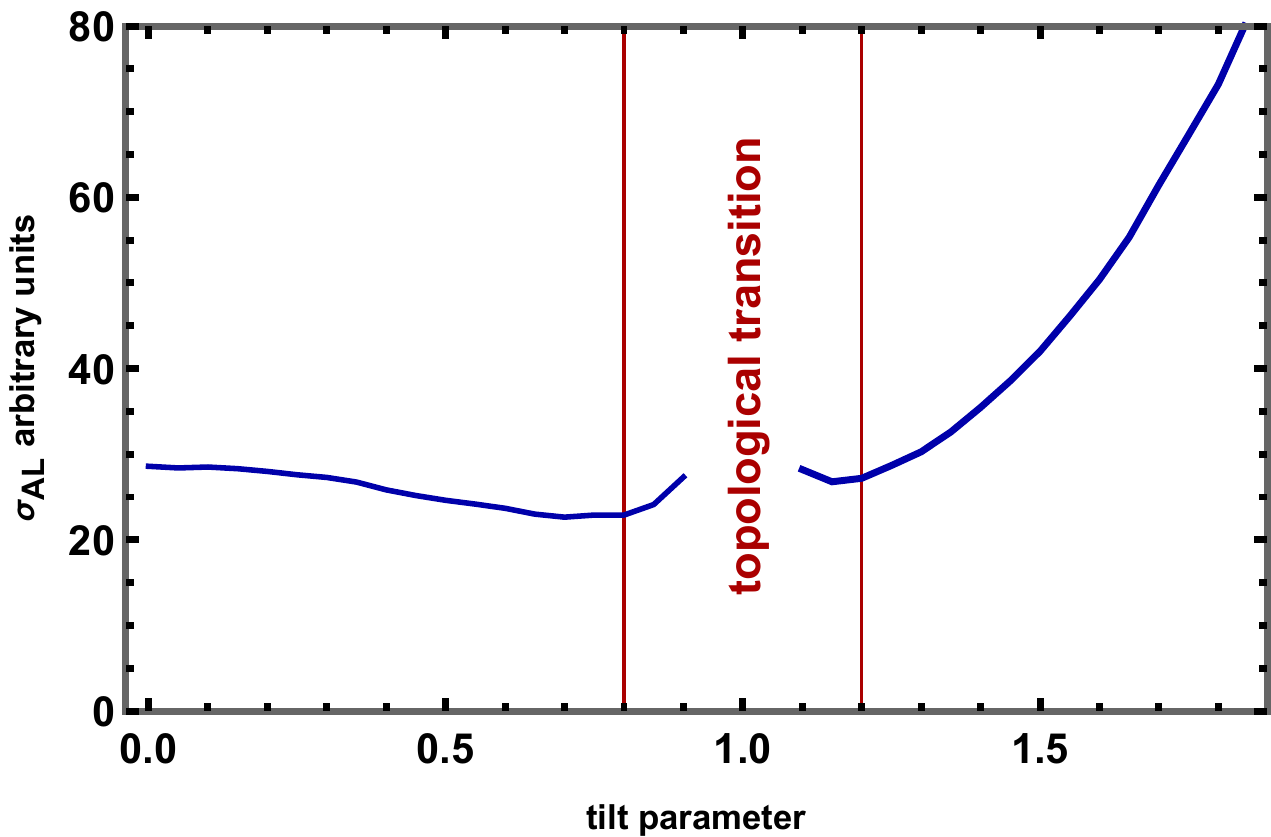}
\caption{ALC of the 3D Weyl superconductors as a function of the tilt cone
parameter $\protect\kappa $}
\label{Fig.2}
\end{figure}

\bigskip

For 2D and 1D cases one obtains from Eq.\ref{sigma} :%
\begin{equation}
\sigma _{\alpha \alpha }^{2D}\left( \kappa \right) =\frac{e^{2}\xi _{\alpha
}\left( \kappa \right) }{16\pi d\xi _{\beta }\left( \kappa \right) \tau
\left( \kappa \right) }\text{; }\sigma _{\alpha \alpha }^{1D}\left( \kappa
\right) =\frac{\pi e^{2}\xi _{\alpha }\left( \kappa \right) }{16S\tau \left(
\kappa \right) ^{3/2}}  \label{2D}
\end{equation}%
and presented in Figs. 3,4

Here $d$ is the film thickness and $S$ is the cross-section of the wire, $%
\alpha \neq \beta $.

\begin{figure}[h]
\centering \includegraphics[width=10cm]{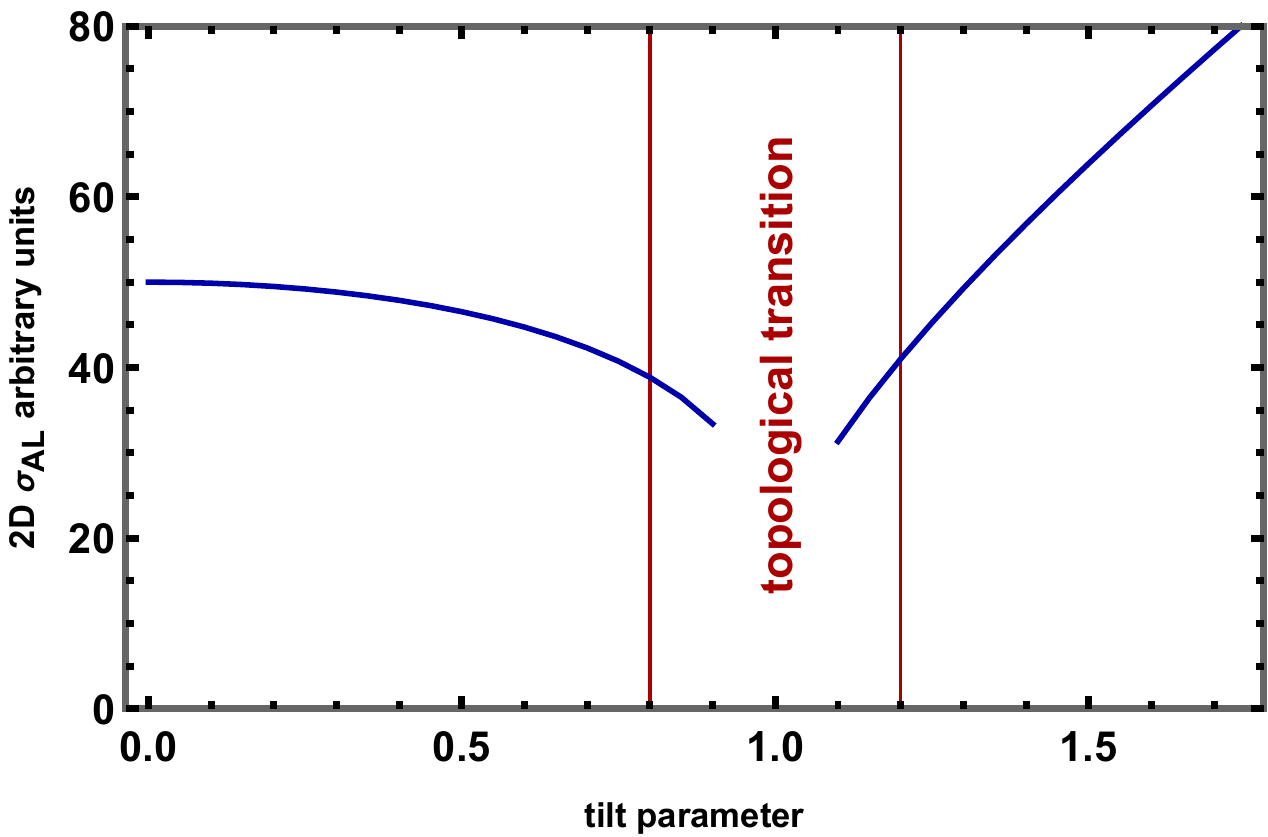}
\caption{ALC of the 2D Weyl superconductors as a function of the tilt cone
parameter $\protect\kappa $}
\label{Fig.3}
\end{figure}

\bigskip 
\begin{figure}[h]
\centering \includegraphics[width=10cm]{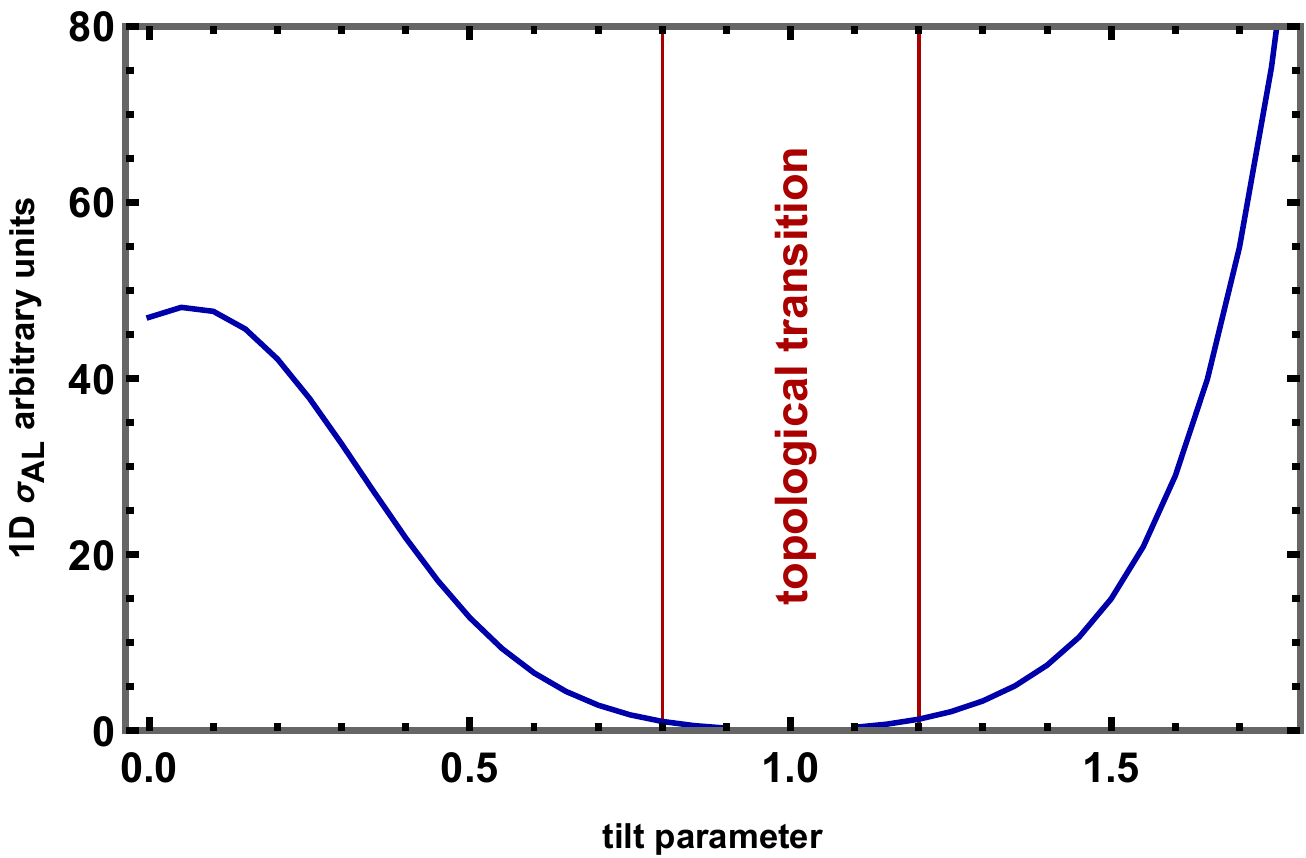}
\caption{ALC of the 1D Weyl superconductors as a function of the tilt cone
parameter $\protect\kappa $}
\label{Fig.4}
\end{figure}

\section{Aslamazov-Larkin conductivity in magnetic field.\textit{\ }}

The ALC in magnetic field was calculated earlier for both for isotropic
system \cite{Terley} and for layered structure with Josephson interaction
between the layers \cite{Varlamov} but these results cannot be used in WSM
with a strong anisotropies of the coherence lengths. In our case with the
external magnetic field directed parallel to the $z-$axis and the
vector-potential $A=\left( Hy,0,0\right) $ , the ac ALC can be calculated
from the dissipative-fluctuation theorem in the Kubo \cite{Kubo} form:

\bigskip 
\begin{equation}
\sigma \left( \omega ,\kappa \right) =\frac{2\pi }{\omega }\tanh \left( 
\frac{\omega }{2T}\right) \int dy\left\langle J_{\omega }\left( 0,\frac{y}{2}%
,0\right) J_{-\omega }\left( 0,-\frac{y}{2},0\right) \right\rangle
\label{sigmamagnetic}
\end{equation}%
Here $J_{\omega }\left( k,\frac{y}{2},k^{\prime }\right) $ is the Fourier
transform of the current density \ref{j} with the fluctuating order
parameter in the form:

\begin{equation}
\Delta \left( \mathbf{r},t\right) =\int dkdp\sum_{n}\Delta _{n,p,k}\left(
t\right) \phi _{n}\left( y-\frac{c}{2eH}k_{x}\right) e^{-ipz-ikx}
\label{deltamag}
\end{equation}

\bigskip

where $\phi _{n,y_{0}}\left( y\right) $ are the normalized eigenfunction of
the electron in magnetic field 
\begin{equation}
\phi _{n,y_{0}}\left( y\right) =\frac{\exp \left( -\left( y-y_{0}\right)
^{2}/2a_{H}^{2}\right) }{\pi ^{1/4}\sqrt{2^{n}a_{H}n!}}H_{n}\left( \frac{%
y-y_{0}}{a_{H}}\right) ;  \label{function}
\end{equation}%
with the eigenvalues 
\begin{equation}
\varepsilon _{n,k_{z}}=\omega _{H}\left( 2n+1\right) +\xi
_{z}^{2}k_{z}^{2},\omega _{H}=\frac{\ eH\sqrt{2}}{c}\xi _{x}\left( \kappa
\right) \xi _{y}\left( \kappa \right) .  \label{epsilon}
\end{equation}

Here $H_{n}$ is the Hermit polynome, $a_{H}^{2}=c\xi _{y}/\sqrt{2}eH\xi
_{x}. $

\bigskip Substituting Eq. (\ref{deltamag}) into Eq.( \ref{j}) one obtains
for the current density after Fourier transform:

\begin{eqnarray}
J_{\omega }\left( 0,\frac{y}{2},0\right)  &=&-2ieD\left( \kappa \right) \xi
_{y}^{2}\Pi ;  \label{Fourier} \\
\Pi  &=&\sum \Pi _{nm}; \\
\Pi _{nm} &=&\int_{-\infty }^{\infty }dkdpd\omega \left[ \Delta
_{n,p,k}\left( \omega +\frac{\Omega }{2}\right) \Delta _{m,p,k}^{\ast
}\left( \omega -\frac{\Omega }{2}\right) F_{n,m}\left( \frac{y}{2}%
,k,p\right) \right] 
\end{eqnarray}

where

\begin{equation}
F_{n,m}\left( y,k_{1},k_{2}\right) =\phi _{n}\left( y-\frac{c}{2eH}%
k_{1}\right) \phi _{m}^{\prime }\left( y-\frac{c}{2eH}k_{2}\right) -\phi
_{m}\left( y-\frac{c}{2eH}k_{2}\right) \phi _{n}^{\prime }\left( y-\frac{c}{%
2eH}k_{1}\right)  \label{F}
\end{equation}

\bigskip

\bigskip Using the random force correlator in the right side of the Eq.(\ref%
{eq3}) one obtains for the correlator of the order parameter $\left\vert
\Delta _{\omega }\right\vert ^{2}=T\left[ D\left( \kappa \right) \left(
\omega ^{2}+E_{n,p,k}^{2}\right) \right] ^{-1},$where $E_{n,p,k}=\varepsilon
_{n,p,k}+\tau .$

\bigskip Substituting this correlator into the Kubo relation for
conductivity and performing the integrations one obtains

for ALC in magnetic field

\bigskip 
\begin{equation}
\sigma _{yy}\left( \kappa ,0\right) =\frac{e^{2}\xi _{y}\left( \kappa
\right) }{16\pi \xi _{x}\left( \kappa \right) \xi _{z}\left( \kappa \right) }%
\sum \left( n+1\right) \left( 
\begin{array}{c}
\frac{1}{\sqrt{\omega _{H}\left( 2n+3\right) +\tau \left( \kappa \right) }}+%
\frac{1}{\sqrt{\omega _{H}\left( 2n+1\right) +\tau \left( \kappa \right) }}-
\\ 
-\frac{2}{\sqrt{\omega _{H}\left( 2n+2\right) +\tau \left( \kappa \right) }}%
\end{array}%
\right)  \label{sigmayy}
\end{equation}

\bigskip

It gives in the limit of a small magnetic fields $H<<H_{c2},$ where $H_{c2}$ 
$=\Phi _{0}\tau \left( \kappa \right) /2\pi \xi _{x}\left( \kappa \right)
\xi _{z}\left( \kappa \right) $ is the upper critical magnetic field, $\Phi
_{0}$is the unit flux%
\begin{equation}
{}\sigma _{yy}\left( \kappa ,0\right) =\frac{e^{2}\xi _{y}\left( \kappa
\right) }{64\pi \xi _{x}\left( \kappa \right) \xi _{z}\left( \kappa \right) }%
\frac{\ 1}{\tau ^{1/2}\left( \kappa \right) }\left\{ 
\begin{array}{c}
\ \ 1-\frac{23}{16}\left( \frac{H}{H_{c2}}\right) ^{2}%
\end{array}%
\right\}  \label{small}
\end{equation}%
In a strong magnetic field ALC diverges close to the $H_{c2}$ . In this case
the main term in the sum is the second ones with \ref{sigmayy} $n=0$\bigskip 
\begin{equation}
\sigma _{yy}\left( \kappa ,0\right) =\frac{e^{2}\xi _{y}\left( \kappa
\right) }{16\pi \xi _{x}\left( \kappa \right) \xi _{z}\left( \kappa \right) }%
\ \sqrt{\frac{H_{c2}^{0}}{H-H_{c2}}}  \label{Hc2}
\end{equation}%
and diverges at the coexistence line.

\section{Discussion and conclusions.}

The Aslamazov-Larkin conductivity in the Weyl metals is described by the
Eqs. \ref{3D},\ref{2D} in the absence of the magnetic field and \ref{sigmayy}%
,\ref{small}, \ref{Hc2} in the presence of the magnetic field \ for
different dimensionality and demonstrate typical for AL conductivity
temperature and field dependences. However, the coherence lengths are the
cone slope dependent (see Figs.2-4 and Fig. 6) demonstrated the main
results. While for 3D and 2D the Aslamazov -Larkin conductivity demonstrate
similar dependence on the tilt parameter (weak variation in Type-I phase and
essential increase in the Type-II phase, the ALC in 1D looks very different
significantly reduces close to the border between Type-I and Type-II phases.

As it was predicted in Ref. \cite{seven}, \cite{eigth}, \cite{nine} the
static pressure can control the cone slope parameter $\kappa $ in Weyl
semimetals. In particular, the type-I WSM realized at large range of the
pressures, while the type-II appears at low pressures. Recently, the
resistance in the WSM was measured in the 3D layered material $HfTe_{5}$ 
\cite{14}.

\begin{figure}[h]
\centering \includegraphics[width=10cm]{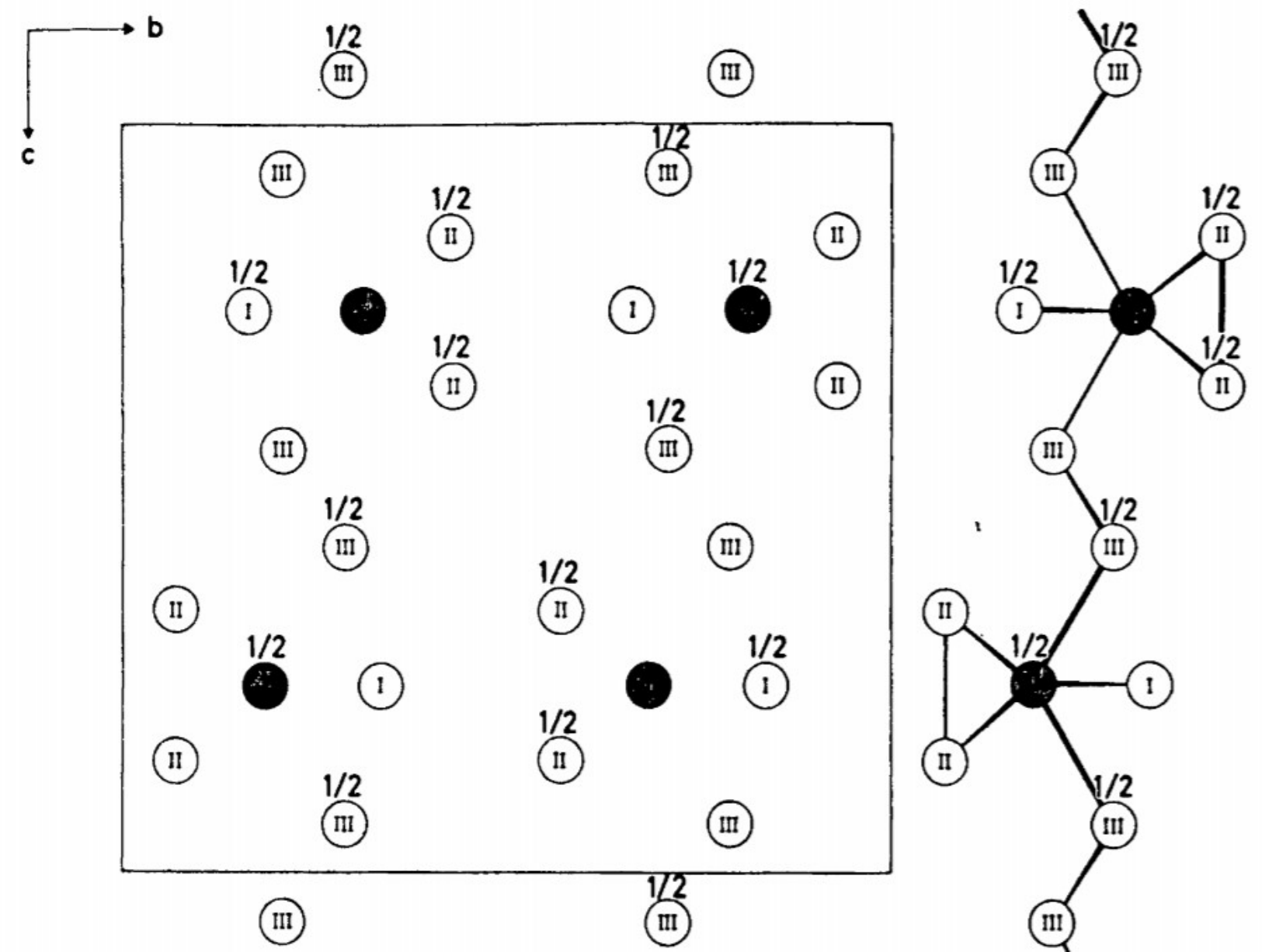}
\caption{ Crystal structure of the $HfTe_{5}$ in (100) projection. Black
balls mark metal atoms, while the rest non-metals. The zig-zag-like layers
run along the b -axis (z axis in our notations). }
\label{Fig.5}
\end{figure}

The crystal structure of $HfTe_{5}$ has been determined by powder X-ray
diffraction experiments \cite{13}. Trigonal prismatic chains of $HfTe_{5}$
run along the a axis, and these prismatic chains are linked via parallel
zigzag chains of Te atoms along the c axis to form a 2D sheet of $HfTe_{5}$
in the ac plane (along the z direction in our notation). The sheets of $%
HfTe_{5}$ stack along the b axis, forming a layered structure \cite{14}
where metallic atoms are surrounded by dielectric as it is shown in Fig. 5 ( 
\cite{31}).

This material shows a well pronounced dependence of the resistivity on the
static pressure. In particular, it was established that below $4.7GPa$ the
type-I WSM is realized while the type- II is realized above $6.1GPa$ . In
the range $4.7-6.1GPa,$ the type-II and type-I Dirac cones coexist. The
critical temperature non-monotonic depending on the pressure allows to
assume that the pressure is directly related to the cone slope $\kappa $
parameter. It is naturally to assume that the Type-II WSM is established at
large pressure and small $\kappa $.

\begin{figure}[h]
\centering \includegraphics[width=10cm]{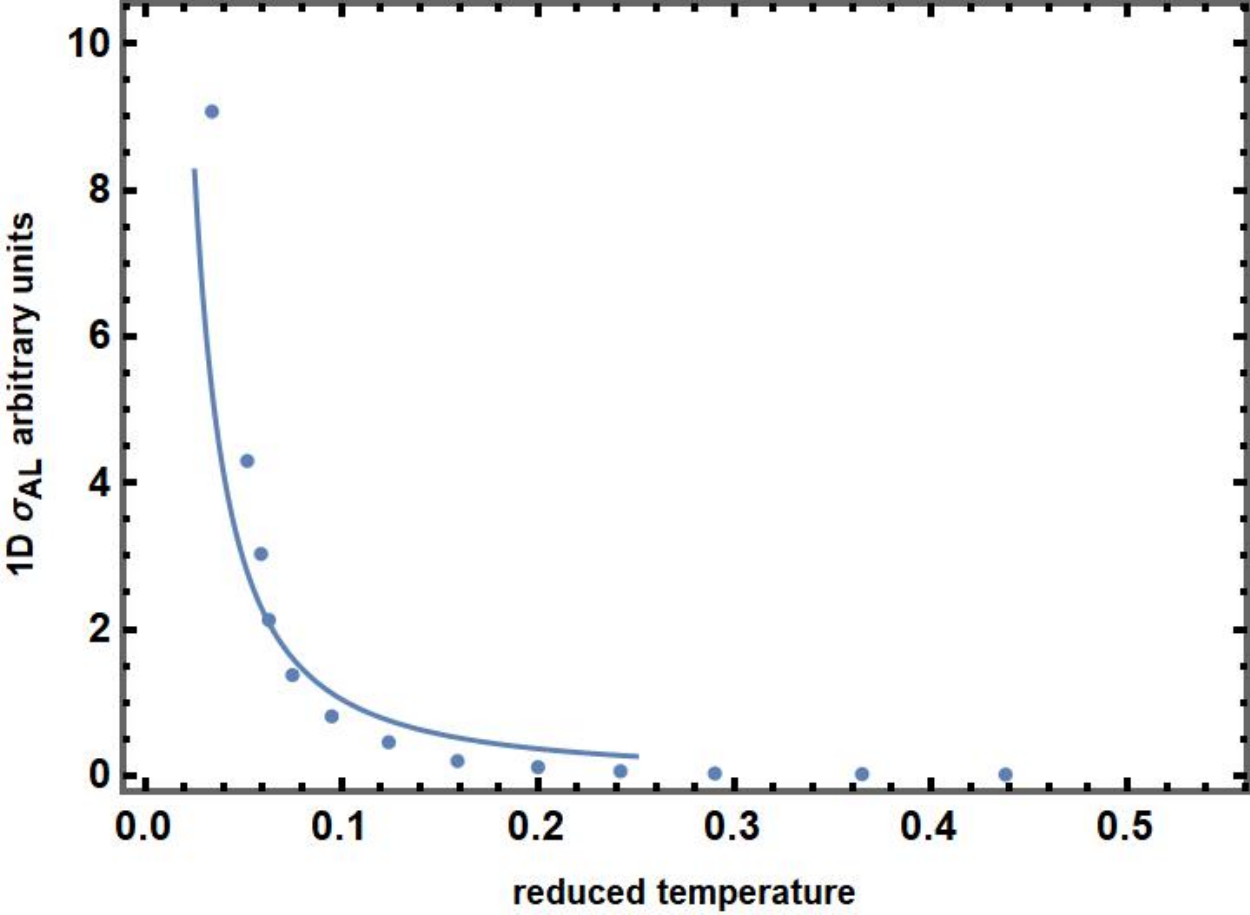}
\caption{ The experimental results (points) and the theoretical fitting by
the (solid curve) for AL conductivity demonstrates the 1D character of the
fluctuations inside the $HfTe_{5}$ layers.}
\label{Fig.6}
\end{figure}

\bigskip

\bigskip

The resistivity was measured in the direction perpendicular to the layers.
The comparison with the theory is presented in Fig.6 and demonstrates a good
agreement with 1D ALC formulae. It shows that the conductivity obeyed the 1D
AL Eq. \ref{2D} (see Fig.4). Measurements of the magnetoresistance when the
current flows perpendicular to the layers (direction) show that its
temperature dependence is well described by the one-dimensional formula for
Aslamazov-Larkin fluctuations. This indicates that the current flows along
the chains of the metal atoms forming one-dimensional channels crossing the
layers.\newpage

$\ $

\textit{Acknowledgements. }

I am grateful Prof. Rosenstein for his attention and assistance. 

\end{document}